\documentclass[aps,prl,showpacs,twocolumn,groupedaddress]{revtex4}  % for submission

\usepackage{graphicx}  % needed for figures
\usepackage{dcolumn}   % needed for some tables
\usepackage{bm}        % for math
\usepackage{amssymb}   % for math
\usepackage{epsfig}

\newcommand{\ttbar}{$t\overline{t}$}

\newcommand{\MCFM}       {{\sc mcfm}}
\newcommand{\PYTHIA}     {{\sc pythia}}
\newcommand{\ALPGEN}       {{\sc alpgen}}

\newcommand{\GEANT}      {{\sc geant}}

\newcommand{\COMPHEP}     {{\sc comphep}}

\newcommand{\rar}       {\rightarrow}
\newcommand{\MET}{$\not\!\!E_T$}

\begin{document}

% the following information is for internal review, please remove them for submission

%\leftline{Version 1.31 as of \today} 

%\leftline{Primary authors: Stephanie Beauceron and Gregorio Bernardi}

\hspace{5.2in} \mbox{Fermilab-Pub-04/288-E}

\title{A Search  for  $Wb\bar{b}$ and $WH$  Production  in $p \bar{p}$ Collisions at $\sqrt{s}=1.96$~TeV}
%\input list_of_authors_r2.tex
% LIST_OF_AUTHORS_R2.TEX                 9/28/04            
%
\author{                                                                      
%% names begin here                                                           
V.M.~Abazov,$^{33}$                                                           
B.~Abbott,$^{70}$                                                             
M.~Abolins,$^{61}$                                                            
B.S.~Acharya,$^{27}$                                                          
M.~Adams,$^{48}$                                                              
T.~Adams,$^{46}$                                                              
M.~Agelou,$^{17}$                                                             
J.-L.~Agram,$^{18}$                                                           
S.H.~Ahn,$^{29}$                                                              
M.~Ahsan,$^{55}$
% ...
\\                                                                            
\vskip 0.30cm                                                                 
\centerline{(D\O\ Collaboration)}                                             
\vskip 0.30cm                                                                 
}                                                                             
\address{                                                                     
\centerline{$^{1}$Universidad de Buenos Aires, Buenos Aires, Argentina}       
\centerline{$^{2}$LAFEX, Centro Brasileiro de Pesquisas F{\'\i}sicas,         
                  Rio de Janeiro, Brazil}                                     
\centerline{$^{3}$Universidade do Estado do Rio de Janeiro,                   
                  Rio de Janeiro, Brazil}                                     
\centerline{$^{4}$Instituto de F\'{\i}sica Te\'orica, Universidade            
                  Estadual Paulista, S\~ao Paulo, Brazil}                     
\centerline{$^{5}$Simon Fraser University, Burnaby, Canada, University of     
                  Alberta, Edmonton, Canada,}                                 
\centerline{McGill University, Montreal, Canada and York University,          
                  Toronto, Canada}                                            
\centerline{$^{6}$Institute of High Energy Physics, Beijing,                  
                  People's Republic of China}                                 
\centerline{$^{7}$Universidad de los Andes, Bogot\'{a}, Colombia}             
\centerline{$^{8}$Charles University, Center for Particle Physics,            
                  Prague, Czech Republic}                                     
\centerline{$^{9}$Czech Technical University, Prague, Czech Republic}         
\centerline{$^{10}$Institute of Physics, Academy of Sciences, Center          
                  for Particle Physics, Prague, Czech Republic}               
\centerline{$^{11}$Universidad San Francisco de Quito, Quito, Ecuador}        
\centerline{$^{12}$Laboratoire de Physique Corpusculaire, IN2P3-CNRS,         
                 Universit\'e Blaise Pascal, Clermont-Ferrand, France}        
\centerline{$^{13}$Laboratoire de Physique Subatomique et de Cosmologie,      
                  IN2P3-CNRS, Universite de Grenoble 1, Grenoble, France}     
\centerline{$^{14}$CPPM, IN2P3-CNRS, Universit\'e de la M\'editerran\'ee,     
                  Marseille, France}                                          
\centerline{$^{15}$Laboratoire de l'Acc\'el\'erateur Lin\'eaire,              
                  IN2P3-CNRS, Orsay, France}                                  
\centerline{$^{16}$LPNHE, Universit\'es Paris VI and VII, IN2P3-CNRS,         
                  Paris, France}                                              
\centerline{$^{17}$DAPNIA/Service de Physique des Particules, CEA, Saclay,    
                  France}                                                     
\centerline{$^{18}$IReS, IN2P3-CNRS, Universit\'e Louis Pasteur, Strasbourg,  
                  France and Universit\'e de Haute Alsace, Mulhouse, France}  
\centerline{$^{19}$Institut de Physique Nucl\'eaire de Lyon, IN2P3-CNRS,      
                   Universit\'e Claude Bernard, Villeurbanne, France}         
\centerline{$^{20}$RWTH Aachen, III. Physikalisches Institut A,               
                   Aachen, Germany}                                           
\centerline{$^{21}$Universit{\"a}t Bonn, Physikalisches Institut,             
                  Bonn, Germany}                                              
\centerline{$^{22}$Universit{\"a}t Freiburg, Physikalisches Institut,         
                  Freiburg, Germany}                                          
\centerline{$^{23}$Universit{\"a}t Mainz, Institut f{\"u}r Physik,            
                  Mainz, Germany}                                             
\centerline{$^{24}$Ludwig-Maximilians-Universit{\"a}t M{\"u}nchen,            
                   M{\"u}nchen, Germany}                                      
\centerline{$^{25}$Fachbereich Physik, University of Wuppertal,               
                   Wuppertal, Germany}                                        
\centerline{$^{26}$Panjab University, Chandigarh, India}                      
\centerline{$^{27}$Tata Institute of Fundamental Research, Mumbai, India}     
\centerline{$^{28}$University College Dublin, Dublin, Ireland}                
\centerline{$^{29}$Korea Detector Laboratory, Korea University,               
                   Seoul, Korea}                                              
\centerline{$^{30}$CINVESTAV, Mexico City, Mexico}                            
\centerline{$^{31}$FOM-Institute NIKHEF and University of                     
                  Amsterdam/NIKHEF, Amsterdam, The Netherlands}               
\centerline{$^{32}$University of Nijmegen/NIKHEF, Nijmegen, The               
                  Netherlands}                                                
\centerline{$^{33}$Joint Institute for Nuclear Research, Dubna, Russia}       
\centerline{$^{34}$Institute for Theoretical and Experimental Physics,        
                  Moscow, Russia}                                             
\centerline{$^{35}$Moscow State University, Moscow, Russia}                   
\centerline{$^{36}$Institute for High Energy Physics, Protvino, Russia}       
\centerline{$^{37}$Petersburg Nuclear Physics Institute,                      
                   St. Petersburg, Russia}                                    
\centerline{$^{38}$Lund University, Lund, Sweden, Royal Institute of          
                   Technology and Stockholm University, Stockholm,            
                   Sweden and}                                                
\centerline{Uppsala University, Uppsala, Sweden}                              
\centerline{$^{39}$Lancaster University, Lancaster, United Kingdom}           
\centerline{$^{40}$Imperial College, London, United Kingdom}                  
\centerline{$^{41}$University of Manchester, Manchester, United Kingdom}      
\centerline{$^{42}$University of Arizona, Tucson, Arizona 85721, USA}         
\centerline{$^{43}$Lawrence Berkeley National Laboratory and University of    
                  California, Berkeley, California 94720, USA}                
\centerline{$^{44}$California State University, Fresno, California 93740, USA}
\centerline{$^{45}$University of California, Riverside, California 92521, USA}
\centerline{$^{46}$Florida State University, Tallahassee, Florida 32306, USA} 
\centerline{$^{47}$Fermi National Accelerator Laboratory, Batavia,            
                   Illinois 60510, USA}                                       
\centerline{$^{48}$University of Illinois at Chicago, Chicago,                
                   Illinois 60607, USA}                                       
\centerline{$^{49}$Northern Illinois University, DeKalb, Illinois 60115, USA} 
\centerline{$^{50}$Northwestern University, Evanston, Illinois 60208, USA}    
\centerline{$^{51}$Indiana University, Bloomington, Indiana 47405, USA}       
\centerline{$^{52}$University of Notre Dame, Notre Dame, Indiana 46556, USA}  
\centerline{$^{53}$Iowa State University, Ames, Iowa 50011, USA}              
\centerline{$^{54}$University of Kansas, Lawrence, Kansas 66045, USA}         
\centerline{$^{55}$Kansas State University, Manhattan, Kansas 66506, USA}     
\centerline{$^{56}$Louisiana Tech University, Ruston, Louisiana 71272, USA}   
\centerline{$^{57}$University of Maryland, College Park, Maryland 20742, USA} 
\centerline{$^{58}$Boston University, Boston, Massachusetts 02215, USA}       
\centerline{$^{59}$Northeastern University, Boston, Massachusetts 02115, USA} 
\centerline{$^{60}$University of Michigan, Ann Arbor, Michigan 48109, USA}    
\centerline{$^{61}$Michigan State University, East Lansing, Michigan 48824,   
                   USA}                                                       
\centerline{$^{62}$University of Mississippi, University, Mississippi 38677,  
                   USA}                                                       
\centerline{$^{63}$University of Nebraska, Lincoln, Nebraska 68588, USA}      
\centerline{$^{64}$Princeton University, Princeton, New Jersey 08544, USA}    
\centerline{$^{65}$Columbia University, New York, New York 10027, USA}        
\centerline{$^{66}$University of Rochester, Rochester, New York 14627, USA}   
\centerline{$^{67}$State University of New York, Stony Brook,                 
                   New York 11794, USA}                                       
\centerline{$^{68}$Brookhaven National Laboratory, Upton, New York 11973, USA}
\centerline{$^{69}$Langston University, Langston, Oklahoma 73050, USA}        
\centerline{$^{70}$University of Oklahoma, Norman, Oklahoma 73019, USA}       
\centerline{$^{71}$Brown University, Providence, Rhode Island 02912, USA}     
\centerline{$^{72}$University of Texas, Arlington, Texas 76019, USA}          
\centerline{$^{73}$Southern Methodist University, Dallas, Texas 75275, USA}   
\centerline{$^{74}$Rice University, Houston, Texas 77005, USA}                
\centerline{$^{75}$University of Virginia, Charlottesville, Virginia 22901,   
                   USA}                                                       
\centerline{$^{76}$University of Washington, Seattle, Washington 98195, USA}  
}                                                                             
%end                                                                          
  % input Dzero author list

\date{\today}

\begin{abstract}
%\begin{center}
%\begin{minipage}{.8\textwidth}
%{\small 
We present a search for  $W b \bar{b}$ production in $p \bar{p}$
collisions at $\sqrt{s}=1.96$~TeV in events containing one electron,
an imbalance in transverse momentum, and two $b$-tagged jets. Using
174~pb$^{-1}$ of integrated luminosity accumulated by the D\O\
experiment at the Fermilab Tevatron collider, and the standard-model 
description of such events, we set a 95\% C.L. upper limit on $W
b \bar{b}$ production of 6.6 pb for $b$ quarks with transverse momenta
$p_T^b > $ 20 GeV and $b \bar{b}$ separation in 
pseudorapidity$-$azimuth space $\Delta {\cal R}_{bb} > 0.75$.  Restricting
the search to optimized $b \bar{b}$ mass intervals provides upper limits
on $WH$ production of 9.0$-$12.2~pb, for Higgs-boson masses
of 105$-$135~GeV.
\end{abstract}
\maketitle

The Higgs boson is the only scalar elementary particle expected in the
standard model (SM). Its discovery would be a major success for the SM
and would provide further insights into the electroweak symmetry
breaking mechanism. The constraints from precision measurements~\cite{EW-fit}
favor a Higgs boson sufficiently light to be accessible at the Fermilab
Tevatron collider. Although the expected 
luminosity necessary
for its discovery
is higher than obtained thus far, the special role of the Higgs boson
in the SM justifies extensive searches for a Higgs-like particle
independent of expected sensitivity. Such studies also provide an 
opportunity to
investigate the main backgrounds,  and in particular
the interesting and thus far unobserved $W b \bar{b}$ production
process.\\
\indent
In this Letter, we present a search for a Higgs ($H$) boson with mass
$m_H$ between 105 and 135 GeV,
in the production channel $p \bar{p} \rar WH \rar e \nu b \bar{b}$, at
$\sqrt{s}=1.96$ TeV.
The expected $WH$ cross section is of the order of 0.2 pb for this mass
range~\cite{higgs}. Our search is based on an integrated luminosity
of 174 $\pm$ 11 pb$^{-1}$ accumulated by the D\O\ experiment during 2002
and 2003.

The experimental signature of $WH \rar e \nu b \bar{b}$ relies
on a final
state with one high $p_T$ electron, two $b$ jets and large imbalance in
transverse momentum (\MET ) resulting from the undetected neutrino. The
dominant backgrounds to $WH$ production are from $W b \bar{b}$, $t \bar{t}$
and single top-quark production. The signal to background ratio is
improved by requiring exactly two jets in the final
state,
because the fraction of $t \bar{t}$ events that contain at most two
reconstructed jets is small.
We use the high statistics $W+\geq 2$ jets data to check 
the validity of
our simulation, but restrict the selection to $W+2$ $b$ jets
for the final results.

The D\O\ detector includes a magnetic tracking system
surrounded by a uranium/liquid-argon calorimeter, which is enclosed
in a muon spectrometer.  The tracking system
 consists of a silicon microstrip
tracker (SMT) and a central fiber tracker (CFT), both located within a
2~T superconducting solenoidal magnet~\cite{run2det}. The SMT and CFT
have
designs optimized for tracking and vertexing capabilities for
pseudorapidities $|\eta|<3$ and  $|\eta|<2.5$, respectively~\cite{foot1}.
The calorimeter has a central section (CC) covering $\eta$ up to $|\eta|
\approx 1.1$, and two end calorimeters (EC) extending coverage to
$|\eta|\approx 4.2$, each housed in a separate
cryostat~\cite{run1det}. For particle identification, the calorimeter
is divided into an electromagnetic (EM) section, followed by fine (FH) and
coarse (CH) hadronic sections.
Scintillators between the CC and EC cryostats provide additional
sampling of developing showers for $1.1<|\eta|<1.4$.  The muon system
consists of a layer of tracking
detectors and scintillation trigger counters in front of 1.8~T
toroids, followed by two similar layers behind the toroids, which
provide muon tracking for $|\eta|<2$.
The luminosity is measured using scintillator arrays located in front 
of the EC cryostats, covering $2.7 < |\eta| < 4.4$. 

Event selection starts with the requirement of an isolated electron,
with $p_T > 20$ GeV, in the central region of $|\eta| <1.1$, but away
from boundaries of calorimeter modules at
periodic azimuthal angle
($\varphi$) values~\cite{calgo}.  Such electrons are required to trigger
the event.  The average trigger efficiency is $(94 \pm 3)\%$ for $W+2$
jets events.
Electron candidates are selected by requiring: (i) at least 90\%
of the energy in a cone of radius $\Delta {\cal R}= \sqrt{(\Delta
\varphi)^2+(\Delta \eta)^2} =0.2$, relative to the shower axis, is
deposited in the EM layers of the calorimeter, i.e., EM fraction {\it
emf} $>$ 0.9; (ii)  isolation, i.e., that the total energy in a
cone of $\Delta {\cal R} < 0.4$ 
centered on the same axis does  not exceed
 the reconstructed electron energy by more
than $10\%$;
(iii) that the energy cluster of the
electron candidate has the characteristics of an EM shower, as
determined by the standard D\O\ shower-shape criteria~\cite{calgo};
(iv) that there is a track pointing to the EM cluster. These four criteria
define the initial electron candidates. Electron selection is  further 
refined
using
an electron likelihood discriminant based on the above estimators, as well as
on additional tracking information.
The combined reconstruction and identification efficiency is
determined from a $Z \rar e^+e^-$ sample to be $(74 \pm
4)\%$ per electron. %(73\% in simulation).

To select $W$ bosons, we require \MET\ $> 25$~GeV. Events with a
second isolated lepton ($e$ or $\mu$~\cite{foot2}) with $p_T >$ 15 GeV
and $|\eta| < 2.4$ are rejected to suppress $Z$~$+$ jets and
$t\bar{t}$ backgrounds.  Only events with a 
primary vertex at $|z| <$ 60 cm
 relative to the center of the detector
are retained.
At least two jets
with $p_T >$ 20 GeV and $|\eta| < 2.5$ are then required.  A jet is
defined as a cluster of  calorimeter towers within a
radius $\Delta {\cal R} =0.5$~\cite{blazey}, having: (i) $ 0.05 <$ {\it
emf} $ < 0.95$; (ii) less than 40\% of its energy in the CH section of
the calorimeter;  (iii) a distance $\Delta {\cal R}$ to any initial electron candidate
greater than 0.5.
The average jet reconstruction and identification efficiency is
$(95 \pm 5)\%$, as determined from $\gamma +$ jet events.
For selecting $b$ jets, we use an
impact-parameter based algorithm~\cite{greder}, which has been
cross-checked with a secondary-vertex reconstruction algorithm.

To improve calorimeter performance, before reconstructing the
calorimeter objects, we use an algorithm that suppresses cells
with negative energy (originating from fluctuations in noise) and
cells with energies four standard  deviations below the average 
electronics noise ($\sigma_n$), when they do not neighbor a
cell of  higher energy, $E > 4 \sigma_n$. 
The EM scale is 
calibrated using the peak in the $Z \rar e^+e^-$ reconstructed mass,
and jet energies are then corrected to the EM scale using $\gamma$+jet
events. These energy corrections, and the transverse momenta of any
muons in the event, are propagated into the calculation of
the \MET , which is 
estimated initially using all (unsuppressed) calorimeter cells.

The D\O\ detector simulation
based on \GEANT ~\cite{GEANT} and the reconstruction and analysis
chain used for data are also used for obtaining 
expectations from the standard model, which 
are normalized to cross sections
measured in data, or to calculations when no such measurements are available.
Small
additional energy smearing in \MET\ and
in the energy of the simulated electrons is used
to obtain better agreement 
between data and simulation. 

Before applying
$b$ tagging, we expect to have two main components in the data: $W+$
jets events and multijet events in which a jet has been misidentified
as an electron (called QCD background in the following).  $W+$ jets
events are simulated using the leading-order matrix-element program
\ALPGEN~\cite{ALPGEN} for the $Wjj$ process
(i.e. production of $W + 2$ partons, which are in our case
gluons or $u,d,s,c$ quarks, 
since the $Wb \bar{b}$ is simulated separately), 
followed by
\PYTHIA~\cite{PYTHIA} for parton showering and hadronization. 
The QCD background is estimated from data 
using  measured probabilities for jets to be
misidentified and accepted
as electrons.

\begin{figure}[t]
\centerline{\psfig{
figure=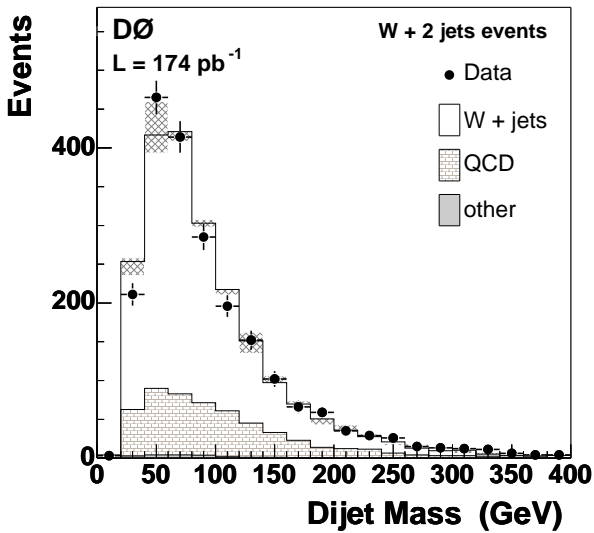,width=3.10in
}}
\caption{Distribution of the dijet invariant mass
of $W+2$ jets events, compared with cumulative contributions
from the QCD background (derived from data), 
the simulation of $W+$jets events and the
other SM backgrounds, which are small before $b$ tagging.
Uncertainties on the simulation from systematics of
the jet energy scale are indicated by the hatched bands.  
The
simulated contributions are  normalized to the integrated luminosity of the data. }
\label {dijet-jes}
\end{figure}

The distribution of the dijet invariant mass in $W+2$ jets events is
shown in Fig.~\ref{dijet-jes}, where it is compared to expectation.
The $Wjj$  expectation is normalized to the data using
the next-to-leading-order (NLO)
\MCFM\ calculation~\cite{MCFM},  providing
a simulated rate for $W+2$ jets events in agreement
with the
measured rate.
Taking into account uncertainties originating from the jet energy
scale, the shape of the distribution is also well described.
The systematic uncertainty associated with the selection of exactly
two jets in the final state has been studied in data and in
simulations. The rates for $W+3$ jets and $W+4$ jets events, after
normalizing to the $W+2$ jets sample, are described by the \ALPGEN\
 and \PYTHIA\ simulation to within $ 15\%$ and $ 6\%$, respectively.  The
resulting systematic uncertainty 
on the expectation 
is $\pm 5\%$.

To search for $W b \bar{b}$ final states and to suppress background,
we apply the $b$-tagging algorithm to jets having at least two tracks,
with $p_T^{track 1 (2)} > 1.0 (0.5)$ GeV.  These
requirements have a typical efficiency per jet of $80\%$ for multijets
events, which is reproduced to within $ 5\%$ by the simulation.  The
$b$-tagging algorithm uses a lifetime probability that is
estimated from the tracks associated with a given jet.  A small probability
corresponds to jets having tracks with large impact parameters that
characterize $b$-hadron decays.  Requiring a probability smaller than
0.7\%, yields a mistag (tagging of $u,d,s$ or gluon jets as $b$ jets)
rate of (0.50$\pm$0.05)\%.  The tagging efficiency for a central
$b$ jet with $p_T$ between 35 and 55 GeV is measured to be
(48$\pm$3)\%.

The tagging efficiency in the simulation is adjusted to
the one measured in data. A study of the $p_T$ and $\eta$ dependence
in data and in simulation indicates a systematic uncertainty on tagging
efficiencies
of $\pm 6\%$. 
When tagging light quarks, there is a larger
systematic uncertainty on the efficiency ($\pm 25\%$)
that originates from the
direct application of the algorithm to simulated events.
This has only a small effect on the final results, since the fraction
of events with two mistagged jets is $<10\%$ of the total
number of $W+2$ $b$-tagged jets.  For the tagging efficiency of $c$ quarks,
 we use the same data/simulation efficiency ratio  as for $b$ quarks.

\begin{figure}[bthp]
\centerline{\psfig{
figure=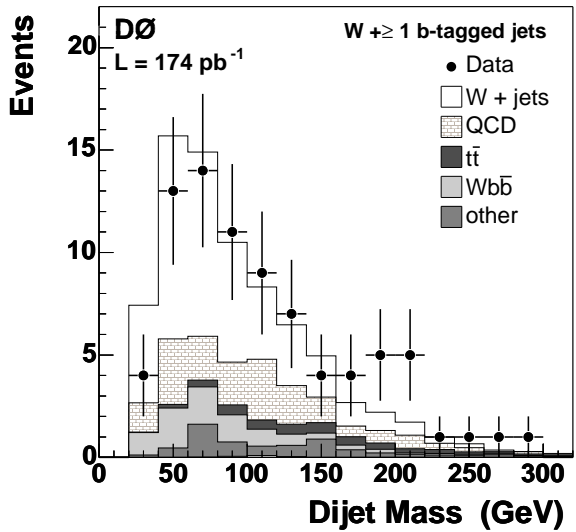,width=3.10in
}}
\caption{
Distribution of the dijet invariant mass for $W+2$ jets events, when
at least one jet is $b$ tagged, compared to expectation
(cumulative). The other SM backgrounds include 
single-top events. 
The
simulated contributions are  normalized to the integrated luminosity of the data. }
\label {dijet-1tag}
\end{figure}

To reduce the
presence of $b$ jets from gluon splitting, and to help assure an
unambiguous determination of  jet flavors in simulation,
we require the separation between the two reconstructed jets
($\Delta {\cal R}$) to be greater than 0.75.
In Fig.~\ref{dijet-1tag} we show the distribution of the dijet mass for
$W+2$ jets events in which at least one jet is $b$ tagged.
The data are well described by the sum of the multijet background and
simulated SM processes (cf. Table~\ref{tab:table3}).  The $t \bar{t}$
contribution is simulated with \PYTHIA\ ($\sigma_{t \bar{t}}=6.77 \pm$
0.42 pb~\cite{ttbar}). Single-top production ($\sigma_{W^*\rar
tb}=1.98 \pm 0.32$ pb, $\sigma_{gW\rar tb}=0.88 \pm 0.13$
pb~\cite{laenen}) is generated with \COMPHEP~\cite{COMPHEP}, assuming
a top-quark mass of 175 GeV, and is shown in Fig.~\ref{dijet-1tag}, in
combination with other processes: $Z \rar ee$, $W \rar \tau \nu$ and
$WZ (\rar b \bar{b}$), which are
simulated using \PYTHIA\ with cross sections of 
255~pb~\cite{cdfWZ}, \ 2775~pb~\cite{cdfWZ},
\ and 0.6 pb~\cite{MCFM}, respectively.
As for the $Wjj$ process,  
the $Wb \bar{b}$ contribution is simulated
using
\ALPGEN\  and \PYTHIA , requiring  $p_T^b> 8$ GeV and $\Delta 
{\cal R}_{bb} >$ 0.4 at the parton level,
with $\sigma_{Wb\bar{b}}$ =3.35 pb computed at
NLO using the \MCFM\ program. 
$WH$ production is
simulated with \PYTHIA\ using the computed cross section at NLO, which
depends on $m_H$~\cite{higgs}.

To further improve  signal/background,
we select events in which a second jet is $b$ tagged.  The
final results for the number of observed and expected events are 
given in Table~\ref{tab:table3}.
Data from the last column are not used 
in the analysis, but provide
 a
check of the accuracy of our expectations 
for  events with two $b$-tagged jets
in the control sample of $W+\geq 3$
jets events, 
which is dominated by $t \bar{t}$ production.

\begin{figure}[htbp]
\centerline{\psfig{
figure=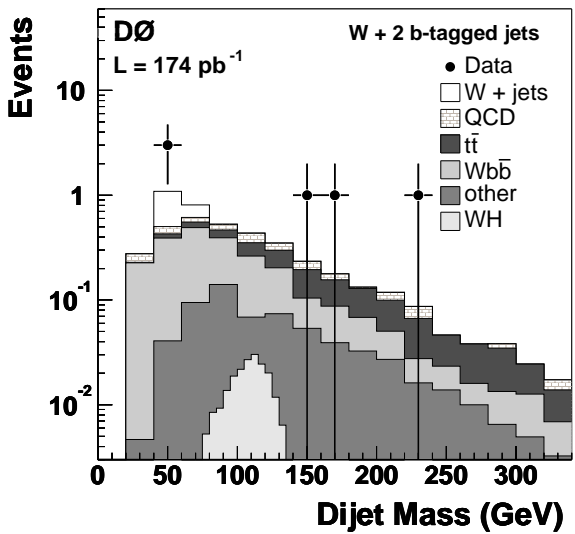,width=3.10in
}
}
\caption{
Distribution of the dijet invariant mass for $W+2~b$-tagged 
events, compared to expectation (cumulative). 
The
simulated contributions are  normalized to the integrated luminosity of the data. 
The expectation 
 for a 115 GeV Higgs boson from $WH$ production is also shown.  }
\label {dijet-2tag}
\end{figure}

The distribution of the dijet mass for events with two $b$-tagged jets
is shown in Fig.~\ref{dijet-2tag}. The expected number of events is
4.4 $\pm 1.2$, of which 1.7 events are expected from $W b \bar{b}$
production.
The dominant systematic uncertainties on the expectation
come from uncertainties on the $b$-tagging 
efficiency (11\%)
and  jet energy corrections.
The uncertainty on the latter 
propagates to uncertainties of 7\% on $W
b \bar{b}$ production, 4\% on single-top and $WH$ production, and 3\%
on $t\bar{t}$ production. The total systematic uncertainty on the
expectation is 26\%, including the uncertainties on cross sections and
luminosity (18\% and 6.5\%, respectively).

\begin{center}
\begin{table*}[htb]
\caption{\label{tab:table3} {
Summary for the $e$+\MET +jets
final state: 
the  numbers of expected  $W+\geq2$ jets and $W+2$ jets events,
before and after $b$ tagging, originating  from 
 $WH$ (for $m_H=115$ GeV), $WZ$,
$Wb \bar{b}$, top production ($t \bar{t}$ and single-top),
QCD multijet background, 
and $W $ or $Z+$jets (excluding $W b \bar{b}$ which is
counted separately) are compared to the numbers of observed events.
The last column shows the same comparison for the control sample of
$W+\geq 3$ jets events that contain two $b$-tagged jets.
}
}
\begin{tabular}{ccccc|c}
\hline
\hline  
& $W + \geq 2$ jets &$W + 2$ jets &$W +$ 2 jets  &$W + 2$ jets &$W + \geq$ 3 jets  \\
          &      &  & (1 $b$-tagged jet)  & (2 $b$-tagged jets) & (2 $b$-tagged jets) \\ 
\hline
$WH$                &   0.6 $\pm$  0.1 &    0.4 $\pm$  0.1 &   0.14 $\pm$ 0.03 &  0.056 $\pm$0.013 &  0.015 $\pm$0.004  \\
$WZ$                &   1.4 $\pm$  0.3 &    1.2 $\pm$  0.3 &   0.38 $\pm$ 0.09 &   0.13 $\pm$ 0.03 &   0.02 $\pm$ 0.01  \\
$Wb\bar{b}$         &  24.7 $\pm$  6.2 &   21.4 $\pm$  5.3 &    6.6 $\pm$  1.5 &   1.72 $\pm$ 0.41 &   0.37 $\pm$ 0.09  \\
\ttbar              &  41.4 $\pm$  8.7 &    8.6 $\pm$  1.8 &    2.7 $\pm$  0.6 &   0.78 $\pm$ 0.19 &   4.63 $\pm$ 1.11  \\
Single-top          &  11.6 $\pm$  2.4 &    8.3 $\pm$  1.7 &    2.7 $\pm$  0.6 &   0.47 $\pm$ 0.11 &   0.30 $\pm$ 0.07  \\
QCD multijet        &   492 $\pm$  108 &    393 $\pm$   86 &   17.1 $\pm$  4.3 &   0.50 $\pm$ 0.20 &   0.92 $\pm$ 0.37  \\
$W$ or $Z$+jets   &  2008 $\pm$  502 &   1672 $\pm$  418 &   43.0 $\pm$ 12.9 &   0.78 $\pm$ 0.22 &   0.86 $\pm$ 0.24  \\
\hline
Total expectation   &  2580 $\pm$  626 &   2106 $\pm$  513 &   72.6 $\pm$ 20.0 &   4.44 $\pm$ 1.17 &   7.12 $\pm$ 1.89   \\
Observed events     &  2540 &   2116 &     76 &      6 &      7  \\
\hline
\hline
\end{tabular}
\end{table*}
\end{center}

\vspace*{-1.2cm}

\vspace*{-0.6cm}

Assuming that the six observed events are consistent with the SM,
without contributions from $W b \bar{b}$ and $WH$, 
and using the $W b \bar{b}$ signal efficiency of (0.90 $\pm$
0.14)\%, we set a 95\% C.L. upper limit of 6.6 pb on the $W b
\bar{b}$ cross section, for $p_T^b > 20$ GeV and $\Delta {\cal R}_{bb}
>$ 0.75~\cite{foot3}. The limits on the cross sections are
obtained using a Bayesian approach~\cite{Bert} that takes 
account of both statistical
and systematic uncertainties.

The expected contribution from the $b \bar{b}$ decay of a SM  Higgs
boson, with $m_H =$ 115 GeV produced with a $W$, 
is also shown in
Fig.~\ref{dijet-2tag}, and amounts to $0.06$ events.  The mean and width
of a Gaussian fit to this expected contribution in the  mass
window 85--135 GeV
are 110 and 16 GeV, a relative resolution of $(14 \pm 1)\%$.  Similar
resolutions are obtained for Higgs-boson masses in the 105-135 GeV region.

No events are observed in the dijet mass window of 85--135 GeV.  The
expected SM background (including $W b \bar{b}$) is 1.07 $\pm$ 0.26
events, and the expected $WH$ signal is 0.049 $\pm$ 0.012 events, with
a signal efficiency of $(0.21 \pm 0.03)\%$.
In the absence of a signal, we set a limit on the cross section for
$\sigma( p\bar{p} \rar WH) \times BR(H \rar b \bar{b}) $ of 9.0 pb at
the 95\% C.L., for a 115 GeV Higgs boson.

\begin{figure}[htbp]
\centerline{\psfig{
figure=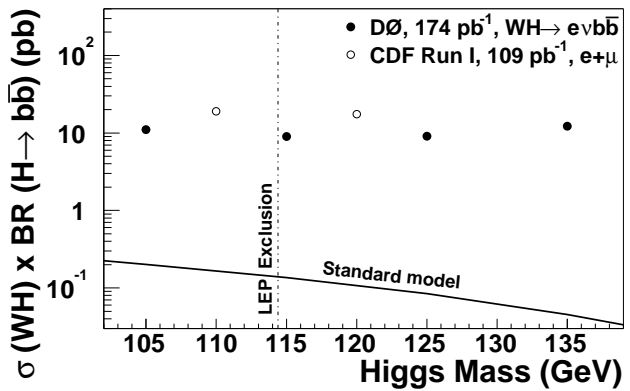,width=3.4in
}
}
\caption{
95\% C.L. upper limit
on   $\sigma(p\bar{p} \rar WH) \times BR(H\rar b\bar{b})$ 
compared  to  the SM expectation at $\sqrt{s}=1.96$ TeV, and to
CDF results~\cite{cdfsearch}, which were obtained at $\sqrt{s}=1.8$ TeV.
The predicted $WH$ cross section
at 1.96 TeV is approximately 15\% larger than that at 1.8 TeV.}
\label {hlimit}
\end{figure}

The same study was performed for $m_H= 105,
125$ and 135 GeV, for which 0, 0 and 1 event were observed in the
corresponding mass windows.  The resulting limits (11.0, 9.1 and 12.2
pb, respectively) are compared to the SM expectation in
Fig.~\ref{hlimit}, and to the results published by the CDF
collaboration, using a smaller integrated luminosity of 109
pb$^{-1}$ at $\sqrt{s}=1.8$ TeV, but 
for combined $e$ and $\mu$ channels~\cite{cdfsearch}.

In conclusion, we have performed a search for the $W b \bar{b}$ final 
state, and have set an upper limit of 6.6 pb on this largest expected
background to $WH$ associated production. We have studied the dijet
mass spectrum of two $b$-tagged jets in the region where we have
the best sensitivity to a SM Higgs boson, and for Higgs-boson masses
between 105 and 135 GeV we set 95\% C.L. upper limits  between 9.0 and
12.2 pb on the  cross section for $WH$ production multiplied by the branching
ratio for $H \rar b \bar{b}$.

We thank the staffs at Fermilab and collaborating institutions, 
and acknowledge support from the 
Department of Energy and National Science Foundation (USA),  
Commissariat  \` a l'Energie Atomique and 
CNRS/Institut National de Physique Nucl\'eaire et 
de Physique des Particules (France), 
Ministry of Education and Science, Agency for Atomic 
   Energy and RF President Grants Program (Russia),
CAPES, CNPq, FAPERJ, FAPESP and FUNDUNESP (Brazil),
Departments of Atomic Energy and Science and Technology (India),
Colciencias (Colombia),
CONACyT (Mexico),
KRF (Korea),
CONICET and UBACyT (Argentina),
The Foundation for Fundamental Research on Matter (The Netherlands),
PPARC (United Kingdom),
Ministry of Education (Czech Republic),
Natural Sciences and Engineering Research Council and 
WestGrid Project (Canada),
BMBF and DFG (Germany),
A.P.~Sloan Foundation,
Research Corporation,
Texas Advanced Research Program,
and the Alexander von Humboldt Foundation.
%
   % input acknowledgement


\begin{thebibliography}{99}
\bibitem{foot1}
The pseudorapidity is defined as a function
of the polar angle $\theta$ as $\eta \equiv - \ln(\tan{\frac{\theta}{2}})$.
\bibitem{foot2} The muon is considered isolated if its distance from
the closest jet is $\Delta {\cal R}>0.4$.
\end{thebibliography}
\end{document}